# Foundations of Newtonian Dynamics: An Axiomatic Approach for the Thinking Student [1]

## C. J. Papachristou [2]

*Department of Physical Sciences, Hellenic Naval Academy, Piraeus 18539, Greece*

**Abstract.** Despite its apparent simplicity, Newtonian mechanics contains conceptual subtleties that may cause some confusion to the deep-thinking student. These subtleties concern fundamental issues such as, e.g., the number of independent laws needed to formulate the theory, or, the distinction between genuine physical laws and derivative theorems. This article attempts to clarify these issues for the benefit of the student by revisiting the foundations of Newtonian dynamics and by proposing a rigorous axiomatic approach to the subject. This theoretical scheme is built upon two fundamental postulates, namely, conservation of momentum and superposition property for interactions. Newton's laws, as well as all familiar theorems of mechanics, are shown to follow from these basic principles.

## 1. Introduction

Teaching introductory mechanics can be a major challenge, especially in a class of students that are not willing to take anything for granted! The problem is that, even some of the most prestigious textbooks on the subject may leave the student with some degree of confusion, which manifests itself in questions like the following:

- Is the law of inertia (Newton's first law) a law of motion (of free bodies) or is it a statement of existence (of inertial reference frames)?
- Are the first two of Newton's laws independent of each other? It appears that the first law is redundant, being no more than a special case of the second law!
- Is the second law a true law or a definition (of force)?
- Is the third law more fundamental than conservation of momentum, or is it the other way around?
- Does the "parallelogram rule" for composition of forces follow trivially from Newton's laws, or is an additional, independent principle required?
- And, finally, what is the minimum number of *independent* laws needed in order to build a complete theoretical basis for mechanics?

In this article we describe an axiomatic approach to introductory mechanics that is both rigorous and pedagogical. It purports to clarify issues like the ones mentioned above, at an early stage of the learning process, thus aiding the student to acquire a deep understanding of the basic ideas of the theory. It is not the purpose of this article, of course, to present an outline of a complete course of mechanics! Rather, we will focus on the most fundamental concepts and principles, those that are taught at the early chapters of dynamics (we will not be concerned with kinematics, since this subject confines itself to a description of motion rather than investigating the physical laws governing this motion).

---

[1] See Note at the end of the article.
[2] papachristou@snd.edu.gr

The axiomatic basis of our approach consists of two fundamental postulates, presented in Section 3. The first postulate (*P1*) embodies both the existence of *inertial reference frames* and the *conservation of momentum*, while the second one (*P2*) expresses a *superposition principle for interactions*. The *law of inertia* is deduced from *P1*.

In Sec. 4, the concept of *force* on a particle subject to interactions is defined (as in *Newton's second law*) and *P2* is used to show that a composite interaction of a particle with others is represented by a vector sum of forces. Then, *P1* and *P2* are used to derive the *action-reaction law*. Finally, a generalization to systems of particles subject to external interactions is made.

For completeness of presentation, certain derivative concepts such as angular momentum, work, kinetic energy, etc., are discussed in Sec. 5. To make the article self-contained, proofs of all theorems are included.

## 2. A critical look at Newton's theory

There have been several attempts to reexamine Newton's laws even since Newton's time. Probably the most important revision of Newton's ideas – and the one on which modern mechanics teaching is based – is that due to Ernst Mach (1838-1916) (for a beautiful discussion of Mach's ideas, see the classic article by H. A. Simon [1]). Our approach differs in several aspects from those of Mach and Simon, although all these approaches share common characteristics in spirit. (For a historical overview of the various viewpoints regarding the theoretical basis of classical mechanics, see, e.g., the first chapter of [2].)

The question of the *independence* of Newton's laws has troubled many generations of physicists. In particular, still on this day some authors assert that the first law (the law of inertia) is but a special case of the second law. The argument goes as follows:

> *"According to the second law, the acceleration of a particle is proportional to the total force acting on it. Now, in the case of a free particle the total force on it is zero. Thus, a free particle must not be accelerating, i.e., its velocity must be constant. But, this is precisely what the law of inertia says!"*

Where is the error in this line of reasoning? Answer: The error rests in regarding the acceleration as an absolute quantity independent of the observer that measures it. As we well know, this is not the case. In particular, the only observer *entitled* to conclude that a non-accelerating object is subject to no net force is an *inertial observer*, one who uses an *inertial frame of reference* for his/her measurements. It is precisely the law of inertia that *defines* inertial frames and *guarantees* their existence. So, without the first law, the second law becomes indeterminate, if not altogether wrong, since it would appear to be valid relative to any observer regardless of his/her state of motion. It may be said that the first law defines the "terrain" within which the second law acquires a meaning. Applying the latter law without taking the former one into account would be like trying to play soccer without possessing a soccer field!

The completeness of Newton's laws is another issue. Let us see a significant example: As is well known, the *principle of conservation of momentum* is a direct consequence of Newton's laws. This principle dictates that the total momentum of a system of particles is constant in time, relative to an inertial frame of reference, when the total external force on the system vanishes (in particular, this is true for an *isolated* system of particles, i.e., a system subject to no external forces). But, when proving this principle we take it for granted that the total force on each particle is the vector sum of all forces (both internal and external) acting on it. This is *not* something that follows trivially from Newton's laws, however! In fact, it was Daniel Bernoulli who first stated this *principle of superposition* after Newton's death. This means that classical Newtonian mechanics is built upon a total of *four* – rather than just three – basic laws.

The question now is: can we somehow "compactify" the axiomatic basis of Newtonian mechanics in order for it to consist of a smaller number of independent principles? At this point it is worth taking a closer look at the principle of conservation of momentum mentioned above. In particular, we note the following:

- For an isolated "system" consisting of a single particle, conservation of momentum reduces to the law of inertia (the momentum, thus also the velocity, of a free particle is constant relative to an inertial frame of reference).
- For an isolated system of two particles, conservation of momentum takes us back to the action-reaction law (Newton's third law).

Thus, starting with four fundamental laws (the three laws of Newton plus the law of superposition) we derived a new principle (conservation of momentum) that yields, as special cases, two of the laws we started with. The idea is then that, by taking *this* principle as our fundamental physical law, the number of independent laws necessary for building the theory would be reduced.

How about Newton's second law? We take the view, adopted by several authors including Mach himself (see, e.g., [1,3-7]) that this "law" should be interpreted as the *definition* of force in terms of the rate of change of momentum.

We thus end up with a theory built upon *two* fundamental principles, i.e., the conservation of momentum and the principle of superposition. In the following sections these ideas are presented in more detail.

## 3. The fundamental postulates and their consequences

We begin with some basic definitions.

*Definition 1.* A *frame of reference* (or *reference frame*) is a system of coordinates (or axes) used by an observer to measure physical quantities such as the position, the velocity, the acceleration, etc., of any particle in space. The position of the observer him/herself is assumed *fixed* relative to his/her own frame of reference.

*Definition 2.* An *isolated system of particles* is a system of particles subject only to their mutual interactions, i.e., subject to no *external* interactions. Any system of particles subject to external interactions that somehow cancel one another in order to make the system's motion identical to that of an isolated system will also be considered "isolated". In particular, an isolated system consisting of a single particle is called a *free particle*.

Our first fundamental postulate of mechanics is stated as follows:

***Postulate 1.*** A class of frames of reference (*inertial frames*) exists such that, for any *isolated* system of particles, a vector equation of the following form is valid:

$$\sum_i m_i \vec{v}_i = \text{constant in time} \tag{1}$$

where $\vec{v}_i$ is the velocity of the particle indexed by $i$ ($i = 1, 2, \cdots$) and where $m_i$ is a constant quantity associated with this particle, which quantity is independent of the number or the nature of interactions the particle is subject to.

We call $m_i$ the *mass* and $\vec{p}_i = m_i \vec{v}_i$ the *momentum* of the *i*th particle. Also, we call

$$\vec{P} = \sum_i m_i \vec{v}_i = \sum_i \vec{p}_i \tag{2}$$

the *total momentum* of the system relative to the considered reference frame. Postulate 1, then, expresses the *principle of conservation of momentum*: the total momentum of an isolated system of particles, relative to an inertial reference frame, is constant in time. (The same is true, in particular, for a free particle.)

***Corollary 1.*** A free particle moves with constant velocity (i.e., with no acceleration) relative to an *inertial* reference frame.

***Corollary 2.*** Any two free particles move with constant velocities relative to each other (their relative velocity is constant and their relative acceleration is zero).

***Corollary 3.*** The position of a free particle may define the origin of an inertial frame of reference.

We note that Corollaries 1 and 2 constitute alternate expressions of the *law of inertia* (*Newton's first law*).
By *inertial observer* we mean an "intelligent" free particle, i.e., one that can perform measurements of physical quantities such as velocity or acceleration. By convention, the observer is assumed to be located at the origin of his/her own inertial frame of reference.

***Corollary 4.*** Inertial observers move with constant velocities (i.e., they do not accelerate) relative to one another.

Consider now an isolated system of two particles of masses $m_1$ and $m_2$. Assume that the particles are allowed to interact for some time interval $\Delta t$. By conservation of momentum relative to an inertial frame of reference, we have:

$$\Delta(\vec{p}_1 + \vec{p}_2) = 0 \;\Rightarrow\; \Delta \vec{p}_1 = -\Delta \vec{p}_2 \;\Rightarrow\; m_1 \Delta \vec{v}_1 = -m_2 \Delta \vec{v}_2 \;.$$

We note that the changes in the velocities of the two particles within the (arbitrary) time interval $\Delta t$ must be in opposite directions, a fact that is verified experimentally. Moreover, these changes are independent of the particular inertial frame used to measure the velocities (although, of course, the velocities themselves *are* frame-dependent!). This latter statement is a consequence of the constancy of the relative velocity of any two inertial observers (the student is invited to explain this in detail). Now, taking magnitudes in the above vector equation, we have:

$$\frac{|\Delta \vec{v}_1|}{|\Delta \vec{v}_2|} = \frac{m_2}{m_1} = \text{constant} \qquad (3)$$

regardless of the kind of interaction or the time $\Delta t$ (which also is an experimentally verified fact). These demonstrate, in practice, the validity of the first postulate. Equation (3) allows us to specify the mass of a particle numerically, relative to the mass of some other particle (which particle may arbitrarily be assigned a unit mass), by letting the two particles interact for some time. As argued above, the result will be independent of the specific inertial frame used by the observer who makes the measurements. That is, in the classical theory, *mass is a frame-independent quantity*.

So far we have examined the case of isolated systems and, in particular, free particles. Consider now a particle subject to interactions with the rest of the world. Then, in general (unless these interactions somehow cancel one another), the particle's momentum will not remain constant relative to an *inertial* reference frame, i.e., will be a function of time. Our second postulate, which expresses the *superposition principle for interactions*, asserts that external interactions act on a particle *independently of one another* and their effects are superimposed.

***Postulate 2.*** If a particle of mass $m$ is subject to interactions with particles $m_1, m_2, \cdots$, then, at each instant $t$, the rate of change of this particle's momentum relative to an inertial reference frame is equal to

$$\frac{d\vec{p}}{dt} = \sum_i \left(\frac{d\vec{p}}{dt}\right)_i \qquad (4)$$

where $(d\vec{p}/dt)_i$ is the rate of change of the particle's momentum due solely to the interaction of this particle with the particle $m_i$ (i.e., the rate of change of $\vec{p}$ *if* the particle $m$ interacted *only* with $m_i$).

## 4. The concept of force and the Third Law

We now *define* the concept of force, in a manner similar to *Newton's second law*:

***Definition 3.*** Consider a particle of mass $m$ that is subject to interactions. Let $\vec{p}(t)$ be the particle's momentum as a function of time, as measured relative to an *inertial* reference frame. The vector quantity

$$\vec{F} = \frac{d\vec{p}}{dt} \qquad (5)$$

is called the *total force* acting on the particle at time $t$.

Taking into account that, for a single particle, $\vec{p} = m\vec{v}$ with fixed $m$, we may rewrite Eq. (5) in the equivalent form,

$$\vec{F} = m\vec{a} = m\frac{d\vec{v}}{dt} \tag{6}$$

where $\vec{a}$ is the particle's acceleration at time $t$. Given that both the mass and the acceleration (prove this!) are independent of the inertial frame used to measure them, we conclude that *the total force on a particle is a frame-independent quantity*.

***Corollary 5.*** Consider a particle of mass $m$ subject to interactions with particles $m_1, m_2, \cdots$. Let $\vec{F}$ be the total force on $m$ at time $t$, and let $\vec{F}_i$ be the force on $m$ due solely to its interaction with $m_i$. Then, by the superposition principle for interactions (Postulate 2) as expressed by Eq. (4), we have:

$$\vec{F} = \sum_i \vec{F}_i \tag{7}$$

***Theorem 1.*** Consider two particles *1* and *2*. Let $\vec{F}_{12}$ be the force on particle *1* due to its interaction with particle *2* at time $t$, and let $\vec{F}_{21}$ be the force on particle *2* due to its interaction with particle *1* at the same instant. Then,

$$\vec{F}_{12} = -\vec{F}_{21} \tag{8}$$

***Proof.*** By the independence of interactions, as expressed by the superposition principle, the forces $\vec{F}_{12}$ and $\vec{F}_{21}$ are independent of the presence or not of other particles in interaction with particles *1* and *2*. Thus, without loss of generality, we may assume that the system of the two particles is isolated. Then, by conservation of momentum and by using Eq. (5),

$$\frac{d}{dt}(\vec{p}_1 + \vec{p}_2) = 0 \;\Rightarrow\; \frac{d\vec{p}_1}{dt} = -\frac{d\vec{p}_2}{dt} \;\Rightarrow\; \vec{F}_{12} = -\vec{F}_{21}.$$

Equation (8) expresses the *action-reaction law* (*Newton's third law*).

***Theorem 2.*** The rate of change of the total momentum $\vec{P}(t)$ of a system of particles, relative to an inertial frame of reference, equals the total *external* force acting on the system at time $t$.

***Proof.*** Consider a system of particles of masses $m_i$ ($i = 1, 2, \cdots$). Let $\vec{F}_i$ be the total *external* force on $m_i$ (due to its interactions with particles *not belonging* to the system), and let $\vec{F}_{ij}$ be the *internal* force on $m_i$ due to its interaction with $m_j$ (by convention, $\vec{F}_{ij} = 0$ when $i=j$). Then, by Eq. (5) and by taking into account Eq. (7),

$$\frac{d\vec{p}_i}{dt} = \vec{F}_i + \sum_j \vec{F}_{ij}.$$

By using Eq. (2) for the total momentum, we have:

$$\frac{d\vec{P}}{dt} = \sum_i \frac{d\vec{p}_i}{dt} = \sum_i \vec{F}_i + \sum_{ij} \vec{F}_{ij}.$$

But,

$$\sum_{ij} \vec{F}_{ij} = \sum_{ji} \vec{F}_{ji} = \frac{1}{2}\sum_{ij}\left(\vec{F}_{ij} + \vec{F}_{ji}\right) = 0,$$

where the action-reaction law (8) has been taken into account. So, finally,

$$\frac{d\vec{P}}{dt} = \sum_i \vec{F}_i = \vec{F}_{ext} \qquad (9)$$

where $\vec{F}_{ext}$ represents the *total external force* on the system.

## 5. Derivative concepts and theorems

Having presented the most fundamental concepts of mechanics, we now turn to some useful derivative concepts and related theorems, such as those of angular momentum and its relation to torque, work and its relation to kinetic energy, and conservative force fields and their association with mechanical-energy conservation.

*Definition 4.* Let $O$ be the origin of an *inertial* reference frame, and let $\vec{r}$ be the position vector of a particle of mass $m$, relative to $O$. The vector quantity

$$\vec{L} = \vec{r} \times \vec{p} = m(\vec{r} \times \vec{v}) \qquad (10)$$

(where $\vec{p} = m\vec{v}$ is the particle's momentum in the considered frame) is called the *angular momentum* of the particle relative to $O$.

*Theorem 3.* The rate of change of the angular momentum of a particle, relative to $O$, is given by

$$\frac{d\vec{L}}{dt} = \vec{r} \times \vec{F} \equiv \vec{T} \qquad (11)$$

where $\vec{F}$ is the *total* force on the particle at time $t$ and where $\vec{T}$ is the *torque* of this force relative to $O$, at this instant.

*Proof.* Equation (11) is easily proven by differentiating Eq. (10) with respect to time and by using Eq. (5).

***Corollary 6.*** If the torque of the total force on a particle, relative to some point $O$, vanishes, then the angular momentum of the particle relative to $O$ is constant in time (*principle of conservation of angular momentum*).

Under appropriate conditions, the above conservation principle can be extended to the more general case of a system of particles (see, e.g., [2-8]).

***Definition 5.*** Consider a particle of mass $m$ in a *force field* $\vec{F}(\vec{r})$, where $\vec{r}$ is the particle's position vector relative to the origin $O$ of an inertial reference frame. Let $C$ be a curve representing the trajectory of the particle from point $A$ to point $B$ in this field. Then, the line integral

$$W_{AB} = \int_A^B \vec{F}(\vec{r}) \cdot d\vec{r} \tag{12}$$

represents the *work* done by the force field on $m$ along the path $C$. (Note: This definition is valid independently of whether or not additional forces, not related to the field, are acting on the particle; i.e., regardless of whether or not $\vec{F}(\vec{r})$ represents the total force on $m$.)

***Theorem 4.*** Let $\vec{F}(\vec{r})$ represent the *total* force on a particle of mass $m$ in a force field. Then, the work done on the particle along a path $C$ from $A$ to $B$ is equal to

$$W_{AB} = \int_A^B \vec{F}(\vec{r}) \cdot d\vec{r} = E_{k,B} - E_{k,A} = \Delta E_k \tag{13}$$

where

$$E_k = \frac{1}{2} m v^2 = \frac{p^2}{2m} \tag{14}$$

is the *kinetic energy* of the particle.

***Proof.*** By using Eq. (6), we have:

$$\vec{F} \cdot d\vec{r} = m \frac{d\vec{v}}{dt} \cdot d\vec{r} = m \vec{v} \cdot d\vec{v} = \frac{1}{2} m d(\vec{v} \cdot \vec{v}) = \frac{1}{2} m d(v^2) = m v dv,$$

from which Eq. (13) follows immediately.

***Definition 6.*** A force field $\vec{F}(\vec{r})$ is said to be *conservative* if a scalar function $E_p(\vec{r})$ (*potential energy*) exists, such that the work on a particle along *any* path from $A$ to $B$ can be written as

$$W_{AB} = \int_A^B \vec{F}(\vec{r}) \cdot d\vec{r} = E_{p,A} - E_{p,B} = -\Delta E_p \tag{15}$$

***Theorem 5.*** If the total force $\vec{F}(\vec{r})$ acting on a particle *m* is conservative, with an associated potential energy $E_p(\vec{r})$, then the quantity

$$E = E_k + E_p = \frac{1}{2}mv^2 + E_p(\vec{r}) \tag{16}$$

(*total mechanical energy* of the particle) remains constant along any path traced by the particle (*conservation of mechanical energy*).

***Proof.*** By combining Eq. (13) (which is generally valid for *any* kind of force) with Eq. (15) (which is valid for *conservative* force fields) we find:

$$\Delta E_k = -\Delta E_p \implies \Delta(E_k + E_p) = 0 \implies E_k + E_p = const.$$

Theorems 4 and 5 are readily extended to the case of a system of particles (see, e.g., [2-8]).

## 6. Some conceptual problems

After establishing our axiomatic basis and demonstrating that the standard Newtonian laws are consistent with it, the development of the rest of mechanics follows familiar paths. Thus, as we saw in the previous section, we can define concepts such as angular momentum, work, kinetic and total mechanical energies, etc., and we can state derivative theorems such as conservation of angular momentum, conservation of mechanical energy, etc. Also, rigid bodies and continuous media can be treated in the usual way [2-8] as systems containing an arbitrarily large number of particles.

Despite the more "economical" axiomatic basis of Newtonian mechanics suggested here, however, certain problems inherent in the classical theory remain. Let us point out a few:

*1. The problem of "inertial frames"*

An inertial frame of reference is only a theoretical abstraction: such a frame cannot exist in reality. As follows from the discussion in Sec. 3, the origin (say, *O*) of an inertial frame coincides with the position of a hypothetical free particle and, moreover, any real free particle moves with constant velocity relative to *O*. However, no such thing as an absolutely free particle may exist in the world. In the first place, every material particle is subject to the infinitely long-range gravitational interaction with the rest of the world. Furthermore, in order for a supposedly inertial observer to measure the velocity of a "free" particle and verify that this particle is not accelerating relative to him/her, the observer must somehow interact with the particle. Thus, no matter how weak this interaction may be, the particle cannot be considered free in the course of the observation.

*2. The problem of simultaneity*

In Sec. 4 we used our two postulates, together with the definition of force, to derive the action-reaction law. Implicit in our arguments was the requirement that action must be *simultaneous* with reaction. As is well known, this hypothesis, which suggests instantaneous action at a distance, ignores the finite speed of propagation of the field associated with the interaction and violates causality.

*3. A dimensionless "observer"*

As we have used this concept, an "observer" is an intelligent free particle capable of making measurements of physical quantities such as velocity or acceleration. Such an observer may use any convenient (preferably rectangular) set of axes (*x*, *y*, *z*) for his/her measurements. Different systems of axes used by this observer have different orientations in space. By convention, the observer is located at the origin *O* of the chosen system of axes.

As we know, inertial observers do not accelerate relative to one another. Thus, the relative velocity of the origins (say, *O* and *O'*) of two different inertial frames of reference is constant in time. But, what if the axes of these frames are in *relative rotation* (although the origins *O* and *O'* move uniformly relative to each other, or even coincide)? How can we tell which observer (if any) is an inertial one?

The answer is that, relative to the system of axes of an inertial frame, a free particle does not accelerate. In particular, relative to a rotating frame, a free particle will appear to possess at least a centripetal acceleration. Such a frame, therefore, cannot be inertial.

As mentioned previously, an object with finite dimensions (e.g., a rigid body) can be treated as an arbitrarily large system of particles. No additional postulates are thus needed in order to study the dynamics of such an object. This allows us to regard momentum and its conservation as more fundamental than angular momentum and its conservation, respectively. In this regard, our approach differs significantly from, e.g., that of Simon [1] who, in his own treatment, places the aforementioned two conservation laws on an equal footing from the outset.

## 7. Summary

Newtonian mechanics is the first subject in Physics an undergraduate student is exposed to. It continues to be important even at the intermediate and advanced levels, despite the predominant role played there by the more general formulations of Lagrangian and Hamiltonian dynamics.

It is this author's experience as a teacher that, despite its apparent simplicity, Newtonian mechanics contains certain conceptual subtleties that may leave the deep-thinking student with some degree of confusion. The average student, of course, is happy with the idea that the whole theory is built upon three rather simple laws attributed to Newton's genius. In the mind of the more demanding student, however, puzzling questions often arise, such as, e.g., how many independent laws we really need to fully formulate the theory, or, which ones should be regarded as truly fundamental laws of Nature, as opposed to others that can be derived as theorems.

This article suggested an axiomatic approach to introductory mechanics, based on two fundamental, empirically verifiable laws; namely, the *principle of conservation of momentum* and the *principle of superposition for interactions*. We showed that all standard ideas of mechanics (including, of course, Newton's laws) naturally follow from these basic principles. To make our formulation as economical as possible, we expressed the first principle in terms of a system of particles and treated the single-particle situation as a special case. To make the article self-contained for the benefit of the student, explicit proofs of all theorems were given.

By no means do we assert, of course, that this particular approach is unique or pedagogically superior to other established methods that adopt different viewpoints regarding the axiomatic basis of classical mechanics. Moreover, as noted in Sec. 6, this approach is not devoid of the usual theoretical problems inherent in Newtonian mechanics (see also [9,10]).

In any case, it looks like classical mechanics remains a subject open to discussion and re-interpretation, and more can always be said about things that are usually taken for granted by most students (this is not exclusively their fault, of course!). Happily, some of my own students do not fall into this category. I appreciate the hard time they enjoy giving me in class!

# References

1. H. A. Simon, *The Axioms of Newtonian Mechanics*, The Philosophical Magazine, Ser. 7, **38** (1947) 888-905.

2. N. C. Rana, P. S. Joag, *Classical Mechanics* (Tata McGraw-Hill, 1991).

3. K. R. Symon, *Mechanics*, 3rd edition (Addison-Wesley, 1971).

4. J. B. Marion, S. T. Thornton, *Classical Dynamics of Particles and Systems*, 4th edition (Saunders College, 1995).

5. M. Alonso, E. J. Finn, *Fundamental University Physics*, Volume I: *Mechanics* (Addison-Wesley, 1967).

6. V. I. Arnold, *Mathematical Methods of Classical Mechanics*, 2nd ed. (Springer-Verlag, 1989).

7. C. J. Papachristou, *Introduction to Mechanics of Particles and Systems* (Hellenic Naval Academy Publications, 2017).[3]

8. H. Goldstein, *Classical Mechanics*, 2nd edition (Addison-Wesley, 1980).

9. C.-E. Khiari, *Newton's Laws of Motion Revisited: Some Epistemological and Didactic Problems*, Lat. Am. J. Phys. Educ. **5** (2011) 10-15.

10. A. E. Chubykalo, A. Espinoza, B. P. Kosyakov, *The Inertial Property of Approximately Inertial Frames of Reference*, Eur. J. Phys. **32** (2011) 1347-1356.

**Note:** This article is a revised version of the following published article:

C. J. Papachristou, *Foundations of Newtonian Dynamics: An Axiomatic Approach for the Thinking Student*, Nausivios Chora Vol. 4 (2012) 153-160;

http://nausivios.snd.edu.gr/docs/partC2012.pdf .

---

[3] Available on the Internet; please search title.